\documentclass[letterpaper]{article} 
\usepackage[]{aaai25}  
\usepackage{times}  
\usepackage{helvet}  
\usepackage{courier}  
\usepackage[hyphens]{url}  
\usepackage{graphicx} 
\usepackage{natbib}  
\usepackage{caption} 
\usepackage{newfloat}
\usepackage{listings}
\DeclareCaptionStyle{ruled}{labelfont=normalfont,labelsep=colon,strut=off} 

\usepackage{xcolor}
\usepackage{xspace}
\usepackage{acronym}
\usepackage{tikz}
\usepackage[listings,skins]{tcolorbox}
\usepackage{booktabs}
\usepackage{amsmath}
\usepackage{subcaption}
\usepackage[capitalise,noabbrev]{cleveref}

\usetikzlibrary{positioning,calc,fit,arrows.meta,shapes,trees,decorations.pathreplacing}

\definecolor{c2}{RGB}{255,176,0}
\definecolor{c1}{rgb}{0.99,0.78,0.07}
\definecolor{c3}{RGB}{220,38,127}
\definecolor{c4}{RGB}{120,94,240}
\definecolor{c5}{RGB}{100,143,255}
\definecolor{c34}{RGB}{170,66,184}

\newcommand{\allnotes}[1]{}
\renewcommand{\allnotes}[1]{#1}

\newcommand{\change}[1]{{\color{blue}#1}}
\newcommand{\changer}[1]{{\color{red}#1}}

\renewcommand{\change}[1]{#1}
\renewcommand{\changer}[1]{#1}

\newcommand{\formhe}[0]{{FormHe}\xspace}

\newcommand{\tikzcircleone}[0]{\tikz[baseline=-0.8ex]{\node[inner sep=0.1em,circle,fill=c2,text=white] {1};}\xspace}
\newcommand{\tikzcircletwo}[0]{\tikz[baseline=-0.8ex]{\node[inner sep=0.1em,circle,fill=c3,text=white] {2};}\xspace}
\newcommand{\tikzcirclethree}[0]{\tikz[baseline=-0.8ex]{\node[inner sep=0.1em,circle,fill=c4,text=white] {3};}\xspace}

\newcommand{\figclingo}[0]{\raisebox{-.6ex}{\includegraphics[width=1em]{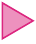}}}

\newcommand{\skchhole}{\begin{tikzpicture}[baseline={(n.base)}] \node[overlay,inner ysep=5pt,inner xsep=5pt,fill=white,rounded corners=2pt] (b) {}; \node[inner sep=1pt,outer sep=0pt] (n) at (b) {\normalfont ?};\end{tikzpicture}}

\acrodef{MCS}{Minimal Correction Subset}
\acrodef{AMCS}{Augmented Minimal Correction Subset}
\acrodef{ACS}{Augmented Correction Subset}
\acrodef{ASP}{Answer Set Programming}
\acrodef{AST}{Abstract Syntax Tree}
\acrodef{AAST}{Anonymized Abstract Syntax Tree}
\acrodef{SBFL}{Spectrum-based Fault Localization}
\acrodef{MBFL}{Mutation-based Fault Localization}
\acrodef{DSL}{Domain Specific Language}
\acrodef{SMT}{Satisfiabi\-li\-ty Modulo Theories}
\acrodef{MSICS}{Minimal Strongly Inconsistent Correction Subset}
\acrodef{LLM}{Large Language Model}
\acrodef{PEFT}{Parameter Efficient Fine Tuning}
\acrodef{LoRa}{Low-Rank Adaptation}
\acrodef{APR}{Automated Program Repair}

\NewTCBListing{asplisting}{ !O{} }{%
  listing only,
  colback=black!10!white,
  colframe=black!15!white,
  hbox,
  leftrule=10pt,
  left=5pt,
  top=-2pt,
  bottom=-3pt,
  listing options={%
      numbers=left,
      basicstyle=\small\ttfamily,
      numberstyle=\color{black}\scriptsize,
      showstringspaces=false,
      breaklines=true,
      escapeinside={\%*}{*)},
      language=prolog,
      aboveskip=\smallskipamount
    },
    #1} 

\newtcbox{\aspinline}{%
  on line,
  boxsep=1pt,
  left=0pt,
  right=0pt,
  top=0pt,
  bottom=0pt,
  colback=black!10!white,
  colframe=black!15!white,
  fontupper={\normalfont\ttfamily\footnotesize}}

\newsavebox\mypostbreak
\savebox\mypostbreak{\raisebox{0ex}[0ex][0ex]{\ensuremath{\color{red}\hookrightarrow\space}}}

\NewEnviron{rcfg}{
  \newcommand{\produces}{&&\rightarrow}
  \newcommand{\alignprod}{& && \phantom{\rightarrow}}
  \newcommand{\term}[1]{\small\texttt{##1}}
  \newcommand{\prodone}[2]{\term{##1(}##2\term{)}}
  \newcommand{\prodtwo}[3]{\term{##1(}##2\term{,}\,##3\term{)}}
  
  \newcommand{\prodfour}[5]{\term{##1(}##2\term{,}\,##3\term{,}\,##4\term{,}\,##5\term{)}}
  \newcommand{\ror}{\enspace|\enspace}
  \begin{alignat*}{2}
    \BODY
  \end{alignat*}
}

\title{Combining Logic with Large Language Models for\\Automatic Debugging and Repair of ASP Programs}
\author{
    Ricardo Brancas\textsuperscript{\rm 1}, Vasco Manquinho\textsuperscript{\rm 1}, Ruben Martins\textsuperscript{\rm 2}
}
\affiliations{
    \textsuperscript{\rm 1}INESC-ID / Instituto Superior Técnico, Lisboa, Portugal\\
    \textsuperscript{\rm 2}Carnegie Mellon University, Pittsburgh, USA\\
    ricardo.brancas@tecnico.ulisboa.pt
}

\begin{document}

\maketitle

\begin{abstract}
Logic programs are a powerful approach for solving NP-Hard problems. However, due to their declarative nature, debugging logic programs poses significant challenges.
Unlike procedural paradigms, which allow for step-by-step inspection of program state, logic programs require reasoning about logical statements for fault localization. This complexity is amplified in learning environments due to students' inexperience.

We introduce \formhe, a novel tool that combines logic-based techniques and Large Language Models to identify and correct issues in Answer Set Programming submissions. \formhe consists of two components: a fault localization module and a program repair module. First, the fault localizer identifies a set of faulty program statements requiring modification. Subsequently, \formhe employs \change{program mutation techniques and Large Language Models} to repair the flawed ASP program. These repairs can then serve as guidance for students to correct their programs.

Our experiments with real buggy programs submitted by students show that \formhe accurately detects faults in 94\% of cases and successfully repairs 58\% of incorrect submissions.
\end{abstract}

\section{Introduction}

Finding bugs can be an arduous task when developing logic programs. This is even more challenging for novice programmers who are still learning. Due to its declarative nature, programmers do not control the flow of execution of a logic program. As such, debugging cannot be done step by step by following the control and data flow of the program. Although some previous attempts to help \ac{ASP} users have been proposed~\cite{DBLP:conf/aaai/Shchekotykhin15}, these assume a high level of interaction~\cite{DBLP:journals/tplp/DodaroGRRS19} and intuition~\cite{DBLP:journals/tplp/OetschPT18} that are not common among novice programmers such as students in a learning context.
Therefore, students primarily use trial and error for debugging since there are no fully automatic tools to help them find the specific parts of the program that are buggy. These hardships can make ASP and declarative programming, in general, difficult to master.

Consider the vertex cover problem. Given a graph $G=(V, E)$ the goal is to find a subset $S$ of vertices ($S \subseteq V$) such that for each edge $(u,v) \in E$ we have that $u \in S$ or $v \in S$
and the cardinality of $S$ is at most $k$.
In ASP, the graph is defined using a \texttt{e/2} predicate, which indicates the existence of an edge between its arguments (vertices of the graph). The student should use the predicate \texttt{sel/1} to indicate which vertices are selected to be in the vertex cover $S$. 
Moreover, a numeric constant, $k$, sets an upper limit on the vertices in set $S$. Consider a graph with 4 edges defined by: \aspinline{e(1,2). e(1,3). e(3,4). e(4,5).}. A possible vertex cover of size 2 for this graph are nodes 1 and 4.

Consider the following \ac{ASP} student submission:
\begin{center}
\begin{minipage}{\linewidth}
\begin{asplisting}
v(X) :- e(X,_).
v(X) :- e(_,X).
%*\textcolor{c2}{k}*) { sel(X) : v(X) } k.
:- not sel(X), not sel(Y), e(X,Y).
\end{asplisting}
\end{minipage}
\end{center}

The first two lines extract the vertex information from the edge predicate, while the third line tells us we want to select \aspinline{k} vertices. 
Finally, the fourth line excludes any solutions where there is an edge between \aspinline{X} and \aspinline{Y}, but neither \aspinline{X} nor \aspinline{Y} are selected.
Note that this program has a bug on the third line: while the problem specification says that \aspinline{k} is an upper bound, the student wrote it so that exactly \aspinline{k} vertices are selected.
Our goal is to automatically identify such bugs and provide students with hints on how to solve them.

We propose \formhe, an automatic tool that combines logic with machine learning to debug and repair ASP programs in a learning environment.
\formhe can perform automatic fault localization on ASP and identify a {\it minimal} set of statements to be corrected or if the program is incomplete (i.e., one or more missing rules). Then, \formhe leverages \change{program mutation techniques} and large language models to automatically repair faulty ASP rule statements and/or generate new rules that are missing from the program. 
The repairs generated by \formhe are then used to provide hints for students to fix their assignments. For the vertex cover example shown above, the following is a possible hint, where ``?'' represents the part that the user should replace:

\begin{center}
\begin{minipage}{\linewidth}
\begin{asplisting}[listing options={%
      numbers=left,
      basicstyle=\normalfont\ttfamily,
      numberstyle=\color{black}\footnotesize,
      showstringspaces=false,
      breaklines=true,
      escapeinside={\%*}{*)},
      language=prolog,
      aboveskip=\smallskipamount,
      firstnumber=3
    }]
%*\setcounter{lstnumber}{3}\skchhole*) { sel(X) : v(X) } k.
\end{asplisting}
\end{minipage}
\end{center}

\section{Primer on ASP}

\acf{ASP}~\cite{DBLP:journals/cacm/BrewkaET11} is a declarative programming language, similar to Prolog~\cite{clocksin2003programming} and Datalog~\cite{DBLP:books/mc/18/MaierTKW18}. 
\ac{ASP} has roots in knowledge representation, uses non-monotonic reasoning, and is inspired by Prolog. The non-monotonic semantics means that adding new premises might decrease the set of inferred facts.

\ac{ASP} programs are comprised of rules.
Consider the rule \aspinline{a :- b, c, not d.}
The left-hand side of a rule is called the \emph{head}, while the right-hand side is the \emph{body}. A head is comprised of an atom, while the body is comprised of a set of literals. A literal is an atom, \aspinline{d} (positive literal), or its negation, \aspinline{not d} (negative literal). This type of negation is called \emph{default negation} and means that literal \aspinline{not d} is assumed to hold unless atom \aspinline{d} is derived to be true.

The intuitive reading of a rule is that if all the positive literals in the body are true and all the negative literals are satisfied,
then the head is also true.
If a rule has no body, it is called a \emph{fact}, and its head is always true.
A rule without the head is called an \emph{integrity constraint}. Integrity constraints filter candidate solutions, meaning the literals in its body must not be jointly satisfied.

Computing a solution of an \ac{ASP} program is done by finding a \emph{stable model} of the formula, called an \emph{answer set}.  
The first step is to \emph{ground} the program, where all the variables in the program are instantiated with specific uses. 
For example, consider the rule \aspinline{b(X) :- a(X).} and the facts \aspinline{a(1). a(2).}. After grounding, this rule would be replaced with the following two rules: \aspinline{b(1) :- a(1). b(2) :- a(2).}
If a program has no negations, then its answer sets can be computed directly from the ground program, similarly to Prolog. Otherwise, the \ac{ASP} solver attempts to find stable models of the formula. The solver will guess an answer set, \(S\), if \(S\) derives itself (it is stable under derivation), then it is an answer set of the formula. We refer to the literature for more details on ASP~\cite{DBLP:series/synthesis/2012Gebser}.

An answer set can also be projected over a set of predicates, \(P\). This projection essentially means removing all predicates not in \(P\) from the answer set. Projected answer sets are helpful when comparing different implementations since they allow us to ignore auxiliary predicates.

\section{Fault Localization} \label{sec:fault-localization}

\formhe focuses on finding and correcting bugs in student assignments. \formhe assumes the availability of one or more \emph{test cases} to test the student submissions. For instance, for the vertex cover problem, these test cases would be the definitions of the graphs to test on. \formhe also requires a reference implementation to check the correction of solutions. An alternative would be to provide all possible answer sets for each test case. However, this is generally harder for large test cases, and teachers already have reference solutions for each exercise in a classroom setting.

Furthermore, \formhe must also know which predicates constitute the solution of a problem (e.g., for the vertex cover example, it would be the \lstinline{sel(V)} predicate).
Note that these predicates are defined in the assignment and do not limit the usability of our approach.
\formhe always projects the answer sets to the solution predicates so that students can define any auxiliary predicates in their programs. 
This section discusses our fault localization methods, which involve identifying minimal strongly inconsistent correction subsets, matching student and reference implementations, and using a large language model-based approach.

\begin{figure}[tb]
    \centering
    \scalebox{0.7}{\begin{tikzpicture}
    [thick,thing/.style={draw=c4,fill=c4!10,inner xsep=0.9em,inner ysep=0.6em},
     component/.style={draw=c2,fill=c2!10,inner xsep=0.9em,inner ysep=0.6em,rounded corners=0.5em}]
    
    \node[component,align=center] (mcs-gen) {\acs{MSICS}\\Generator};
    \node[component] (line-sbfl) [above=1em of mcs-gen,align=center] {Line Matching /\\LLM Classifier};
    \node[component,minimum height=3em] (verifier) [below=3.5em of mcs-gen] {Verifier};

    \node[align=center] (lines) [right=4.5em of line-sbfl.center] {Suspicious\\Lines};
    \node[] (mcss) at (lines |- mcs-gen.east) {\acsp{MSICS}};
    
    \coordinate (verifier1) at ($(verifier.south west)!1/4!(verifier.north west)$);
    \coordinate (verifier2) at ($(verifier.south west)!2/4!(verifier.north west)$);
    \coordinate (verifier3) at ($(verifier.south west)!3/4!(verifier.north west)$);


    \node[isosceles triangle,draw=c3,fill=c3!60,isosceles triangle apex angle=60,minimum height=5pt,minimum width=5pt,inner sep=0] (clingo2) [below left=5pt of verifier.north east] {};
    \node[isosceles triangle,draw=c3,fill=c3!60,isosceles triangle apex angle=60,minimum height=5pt,minimum width=5pt,inner sep=0] (clingo3) [below left=5pt of mcs-gen.north east] {};


    \node[thing,align=left] (acs) [right=6em of $(lines)!.5!(mcss)$] {\acsp*{ACS}};
    \node[thing,align=left] (no-mcs) at (acs |- verifier) {No Faults\\Detected};

    \coordinate (line-sbfl1) at ($(line-sbfl.south west)!1/3!(line-sbfl.north west)$);
    \coordinate (line-sbfl2) at ($(line-sbfl.south west)!2/3!(line-sbfl.north west)$);

    \node[align=right] (stu) [left=7.5em of mcs-gen.center] {Student\\Submission};
    \node[align=right,anchor=east] (ref) at (line-sbfl2 -| stu.east) {Reference\\Implementation};
    \node[anchor=east] (inputs) at (verifier1 -| stu.east) {Test Cases};

    \draw[arrows = {-Straight Barb[length=0.4em]},thick,rounded corners=4pt] (stu.east) -| ($(stu.east)!0.7!(line-sbfl1)$) |- (line-sbfl1);
    \draw[arrows = {-Straight Barb[length=0.4em]},thick,rounded corners=4pt] (stu.east) --  (mcs-gen.west);


    \draw[arrows = {-Straight Barb[length=0.4em]},thick,rounded corners=4pt] (ref.east) -- (line-sbfl2);

    \draw[arrows = {-Straight Barb[length=0.4em]},thick,rounded corners=4pt] (ref.east) -| ($(ref.east)!0.15!(verifier3)$) |- (verifier2);
    \draw[arrows = {-Straight Barb[length=0.4em]},thick,rounded corners=4pt] (stu.east) -| ($(stu.east)!0.46!(verifier2)$) |- (verifier3);
    \draw[arrows = {-Straight Barb[length=0.4em]},thick,rounded corners=4pt] (inputs) -- (verifier1);



    \draw[arrows = {-Straight Barb[length=0.4em]},thick] (verifier) -- (mcs-gen) node[pos=.5,right,align=left] {Missing/Extra\\Answer Sets};

    \draw[] (line-sbfl) -- (lines);
    \draw[] (mcs-gen) -- (mcss);

    \draw[arrows = {Straight Barb[length=0.4em]-},rounded corners=4pt] (acs.west) -- ++(-2.5em,0) |- (lines.east);
    \draw[arrows = {Straight Barb[length=0.4em]-},rounded corners=4pt] (acs.west) -- ++(-2.5em,0) |- (mcss.east);

    \draw[arrows = {-Straight Barb[length=0.4em]},thick] (verifier) -- (no-mcs);

\end{tikzpicture}}
    \caption{System diagram of \formhe's fault localization module. The \figclingo{} represents usage of the Clingo \ac{ASP} solver.}
    \label{fig:fl}
\end{figure}
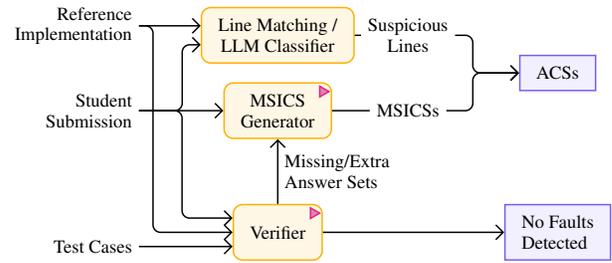

\subsection{Verification}

The overall architecture of \formhe's fault localization module is shown in \cref{fig:fl}.
The first step is verifying if the student submission is correct. A submission is considered correct if (1) all the answer sets it produces are answer sets of the reference implementation and (2) if the reference implementation generates at least one answer set, the submission must also generate at least one answer set. This correctness definition is flexible enough to allow students to add symmetry-breaking constraints~\cite{drescher2011symmetry}.
However, note that the reference implementation must not use any additional constraints that eliminate solutions so that it is ``compatible'' with all correct student implementations, i.e., a valid answer set for the student implementation must be an answer set of the reference. 

If a submission is deemed incorrect, we get a set of \emph{extra} answer sets (answer sets of the student submissions that are not answer sets of the reference implementation) and a set of \emph{missing} answer sets (answer sets of the reference implementation that are not answer sets of the student submission). Using this information, we can identify which lines of the student submission are faulty using \acp{MSICS}.

\subsection{Identifying Minimal Strongly Inconsistent Correction Subsets} \label{sec:mcs}

A subformula \(\phi_s \subseteq \phi\) is strongly inconsistent if \(\phi_s\) is inconsistent and all the supersets \(\phi'\) of the subformula  (\(\phi_s \subseteq \phi' \subseteq \phi\)) are also inconsistent. Given an \emph{inconsistent} formula $\phi = \phi_h \land \phi_s$ where $\phi_h$ is a set of hard constraints and $\phi_s$ is a set of soft constraints, a \acf{MSICS} $\phi_c$ is a minimal set of soft constraints ($\phi_c \subseteq \phi_s$) that need to be removed so that the remaining soft and hard constraints are \emph{not strongly inconsistent}.
For the purpose of fault localization in an ASP program, an \ac{MSICS} is a set of lines of the faulty program that must be removed or modified because it is preventing the program from behaving correctly for a given test case.

Consider the faulty student submission for the vertex cover problem presented in the introduction. \formhe found a test case for which this submission fails:

\begin{center}
\begin{minipage}{.6\linewidth}
\centering
\begin{asplisting}
#const k = 3.
e(1,2).
\end{asplisting}
\end{minipage}%
\begin{minipage}{.4\linewidth}
\centering
\includegraphics[width=.5\linewidth]{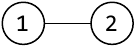}
\end{minipage}
\end{center}

This test case specifies a vertex cover with a maximum size of three for the simple graph shown. The student implementation fails because it uses \texttt{k} as both an upper and lower bound for the size of the cover when it should be just an upper bound.
Hence, the student submission does not output any answer set for this test case. One of the missing answer sets is \aspinline{sel(2)}, since selecting just vertex 2 is a cover of the graph. Using this information, we can compute an \ac{MSICS} of the program.
First, we turn each of the lines of the student program into soft constraints. Next, we specify the test case and the missing answer set as hard constraints \change{(more details in the appendix)}. Finally, using an \ac{MSICS} algorithm \cite{DBLP:conf/sat/Mencia020}, we compute a minimal set of lines that need to be removed or modified such that the answer set can become a solution for the test case.

Due to the non-monotonicity of \ac{ASP}, the \ac{MSICS} approach can sometimes fail to identify all the faulty lines in the program. To overcome this, \formhe can combine the information from the \ac{MSICS} with supplemental sources of information that can select other suspicious lines. Next, we introduce two alternative fault localization methods.

\subsection{Large Language Models}
Deep Learning models have proven to be a powerful tool for fault localization in different domains~\cite{DBLP:conf/issta/LiLZZ19,DBLP:conf/icse/MengW00022,DBLP:conf/icse/YangGMH24}. One way to use \acp{LLM} for fault localization is to transform the model into a classifier by replacing the last layer with a classification head. This approach allows us to generate a score for each line, indicating the likelihood that the line is faulty. While this approach has the downside of requiring that the maximum number of lines in the program be defined at training time, this is not an issue for the small introductory \ac{ASP} problems we are targeting.

Our prompt template contains the problem name (e.g., graph k-coloring), the reference implementation and the student submission. The model has been fine-tuned to receive the prompt and output a score between 0 and 1 for each line. Lines with a score \(\ge 0.5\) are considered faulty.

\subsection{Line Matching}

The Line Matching method finds lines in the student submission that are very similar to lines in the reference implementation yet have small differences. The intuition is that if such lines exist, they are likely to be bugs. On the contrary, very different lines are likely to be due to entirely different implementation approaches instead.

To find these pairs of matching lines, \formhe transforms each line in the submission and the reference implementation into a bag of anonymized nodes.
This transformation ignores the names used in the predicates and variables, as well as the order of most elements. This is important because the order is irrelevant for many syntax elements of ASP programs. After computing the bag of nodes for all lines in both programs, we compute the symmetric difference for each pair of lines from one program to the other. This metric gives us the nodes that need to be removed and/or added to transform one line into another. Then, we create a bipartite graph between the lines of the submission and the lines of the reference implementation, where the weight of each edge is the previously computed distance metric. Finally, we use a perfect matching algorithm \changer{\cite{DBLP:journals/taes/Crouse16}} to find a minimum cost pairing between the lines of the submission and reference implementation. Matched lines from the student submission with a small but non-empty symmetric difference (by default \(\leq 3\)) are reported as likely to be buggy.

\subsection{Choosing a correct implementation}
After \formhe has been used in a class, we obtain several correct solutions for each exercise that are potentially different from the reference implementation. These other correct solutions can be exploited to improve the quality of the similarity-based fault localization methods~\cite{DBLP:conf/pldi/GulwaniRZ18}: the \ac{LLM} approach and the line matching approach.
To achieve this, we use the core technique of the line matching algorithm to compute a distance metric (the total value of the matching) between the student submission and each of the available correct implementations for that problem. Then, we select the lowest distance correct implementation and use it in place of the reference in the similarity-based fault localization methods, improving their performance.

\subsection{Combining Fault Localization Methods} \label{sec:fl-combo}

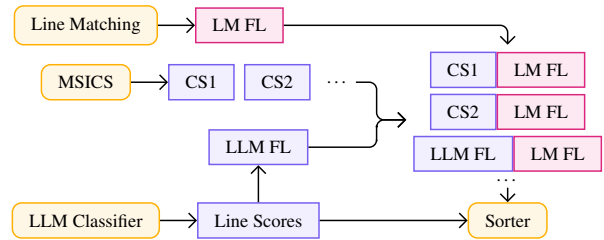
\begin{figure}[tb]
    \centering
    \scalebox{0.7}{\begin{tikzpicture}[thick,thing/.style={draw=c4,fill=c4!10,inner xsep=0.9em,inner ysep=0.6em},
    component/.style={draw=c2,fill=c2!10,inner xsep=0.9em,inner ysep=0.6em,rounded corners=0.5em},
    thing2/.style={draw=c3,fill=c3!10,inner xsep=0.9em,inner ysep=0.6em}]
    
    \node[component] (mcs-strict) {\acs*{MSICS}};
    \node[component] (llm) [below=5.5em of mcs-strict] {\acs*{LLM} Classifier};

    \node[thing] (strict1) [right =2em of mcs-strict] {CS1};
    \node[thing] (strict2) [right =0.5em of strict1] {CS2};
    \node[] (strict3) [right =0.5em of strict2] {\dots};




    \node[thing] (llm-scores) [right =2em of llm] {Line Scores};
    \node[thing] (llm-fl) [above =2em of llm-scores] {LLM FL};

    \node[thing] (fl1) [above right = -0.75em and  4em of strict3] {CS1};
    \node[thing] (fl2) [below = 0.3em of fl1] {CS2};
    \node[thing] (fl5) [below = 0.3em of fl2] {LLM FL};

    \node[component] (line-match) [above=1em of mcs-strict] {Line Matching};

    \node[thing2] (lm-fl) [right=2em of line-match] {LM FL};

    \node[thing2] (lm1) [right = 0em of fl1] {LM FL};
    \node[thing2] (lm2) [right = 0em of fl2] {LM FL};
    \node[thing2] (lm5) [right = 0em of fl5] {LM FL};

    \draw[arrows = {-Straight Barb[length=0.4em]},thick] (line-match) -- (lm-fl);
    \draw[arrows = {-Straight Barb[length=0.4em]},thick] (mcs-strict) -- (strict1);
    \draw[arrows = {-Straight Barb[length=0.4em]},thick] (llm) -- (llm-scores);
    \draw[arrows = {-Straight Barb[length=0.4em]},thick] (llm-scores) -- (llm-fl);

    \coordinate (dots-coord) at ($(fl5.west)!0.5!(lm5.east)$);
    \node[] (fl6) [below = 1em of dots-coord] {\dots};

    \node[fit= (fl1) (lm5) (fl5) (fl6)] (fls) {};

    \draw[arrows = {-Straight Barb[length=0.4em]},thick,rounded corners=4pt] (strict3) -| ($(strict3)!0.566!(fls.west)$) |- (fls.west);
    \draw[arrows = {-Straight Barb[length=0.4em]},thick,rounded corners=4pt] (llm-fl) -| ($(llm-fl)!0.799!(fls.west)-(0,0.42em)$) |- (fls.west);

    \draw[arrows = {-Straight Barb[length=0.4em]},thick,rounded corners=4pt] (lm-fl) -| (fls.north);

    \node[component] (sorter) at (llm-scores -| fls) {Sorter};

    \draw[arrows = {-Straight Barb[length=0.4em]},thick] (fls.south)+(0,.4em) -- (sorter);
    \draw[arrows = {-Straight Barb[length=0.4em]},thick] (llm-scores) -- (sorter.west);

\end{tikzpicture}}
    \caption{Overview of how \formhe combines different fault localization techniques.}
    \label{fig:llm-combo}
\end{figure}

To produce a more robust fault localizer, we combine the different fault localization techniques presented in this section. \Cref{fig:llm-combo} shows an overview of our combination and sorting method.
First, we collect all the \acp{MSICS} produced (CS1, CS2, ...). Then, we use the LLM fault localization approach to obtain another set of faulty lines (LLM FL). For the rest of the process, this set is treated as if it were an \ac{MSICS}.

Next, we compute the set of suspicious lines using the Line Matching approach. These lines are appended to all the \acp{MSICS} (plus the LLM FL), resulting in \acp{ACS}. Finally, we use the Line Scores from the LLM Classifier to sort the \acp{ACS}, in order to choose which one to present to the user and proceed with for the repair module. Each \ac{ACS} is ranked based on the sum of the scores of the lines it contains, with \acp{ACS} with higher scores being ranked first.

\section{Program Repair} \label{sec:repair}

\begin{figure}[tb]
    \centering
    \scalebox{.7}{\begin{tikzpicture}
    [thing/.style={draw=c4,fill=c4!10,inner xsep=0.9em,inner ysep=0.6em},
     component/.style={draw=c2,fill=c2!10,inner xsep=0.9em,inner ysep=0.6em,rounded corners=0.5em}]

    \node[component] (llm) [align=center] {LLM\\Repair};
    \node[component] (enum) [right=of llm,align=center] {Mutation\\Enumerator};

    \coordinate (rep-center) at ($(llm)!1/2!(enum)$);
    
    \node[component,minimum width=12em] (verif) [above=6.5em of rep-center] {Verifier};

    \node[isosceles triangle,draw=c3,fill=c3!60,isosceles triangle apex angle=60,minimum height=5pt,minimum width=5pt,inner sep=0] (clingo3) [below left=5pt of verif.north east] {};

    \coordinate (enumt1) at ($(enum.north)-(0em,0)$);
    \coordinate (enumt2) at ($(enum.north)+(.75em,0)$);

    \coordinate (llmt1) at ($(llm.north)-(0em,0)$);
    \coordinate (llmt2) at ($(llm.north)+(.75em,0)$);

    \coordinate (verif1) at (verif.south -| enumt1);
    \coordinate (verif2) at (verif.south -| enumt2);

    \coordinate (verif3) at (verif.south -| llmt1);
    \coordinate (verif4) at (verif.south -| llmt2);
    

    \draw[arrows = {-Straight Barb[length=0.4em]},thick] (enumt1) -- (verif1) node [pos=.33,left,align=right] (rep-can) {Repair\\Candidate};
    \draw[arrows = {-Straight Barb[length=0.4em]},thick] (verif2) -- (enumt2) node [pos=.33,right,align=left] (rep-fail) {\textit{Failed}};

    \draw[arrows = {-Straight Barb[length=0.4em]},thick] (llmt1) -- (verif3) node [pos=.33,left,align=right] (rep-can2) {Repair\\Candidate};
    \draw[arrows = {-Straight Barb[length=0.4em]},thick] (verif4) -- (llmt2) node [pos=.33,right,align=left] (rep-fail2) {\textit{Failed}};

    

    \node[thing] (repair) [right=9em of verif.center,align=left] {Repaired\\Program};
    \node[thing,anchor=west,align=left] (acs-exhaust) at (repair.west |- enum) {Enumerator\\Exhausted};


    \coordinate (verifa) at ($(verif.south west)!1/3!(verif.north west)$);
    \coordinate (verifb) at ($(verif.south west)!2/3!(verif.north west)$);
    
    \node[align=right] (acs) [left=6em of llm.center] {Student\\Submission};
    \node[align=right] (stu) [below left=1.5em and 6em of llm.center] {ACS};

    \node[align=right] (ref) [below left=1.2em and 10em of verif.center] {Reference\\Implementation};
    \node[align=right,anchor=east] (inputs) at (verifb -| ref.east) {Test\\Cases};

    \coordinate (enuma) at ($(llm.south west)!1/4!(llm.north west)$);
    \coordinate (enumb) at ($(llm.south west)!2/4!(llm.north west)$);
    \coordinate (enumc) at ($(llm.south west)!3/4!(llm.north west)$);
    
    \draw[arrows = {-Straight Barb[length=0.4em]},thick,rounded corners=3pt] (acs.east) -- (enumb);
    \draw[arrows = {-Straight Barb[length=0.4em]},thick] (enum) -- (acs-exhaust);
    \draw[arrows = {-Straight Barb[length=0.4em]},thick,rounded corners=3pt] (stu.east) -| ($(stu.east)!0.3!(enuma)$) |- (enuma);
    \draw[arrows = {-Straight Barb[length=0.4em]},thick,rounded corners=3pt] (ref.east) -| ($(ref.east)!0.26!(enumc)$) |- (enumc);

    \draw[arrows = {-Straight Barb[length=0.4em]},thick,rounded corners=3pt] (ref.east) -| ($(ref.east)!0.25!(verifa)$) |- (verifa);
    \draw[arrows = {-Straight Barb[length=0.4em]},thick,rounded corners=3pt] (inputs.east) -- (verifb);

    \draw[arrows = {-Straight Barb[length=0.4em]},thick] (verif) -- (repair);
    \draw[arrows = {-Straight Barb[length=0.4em]},thick] (llm) -- (enum);

\end{tikzpicture}}
    \caption{System diagram of \formhe's repair module. The \figclingo{} symbol represents usage of the Clingo \ac{ASP} solver.}
    \label{fig:repair}
\end{figure}
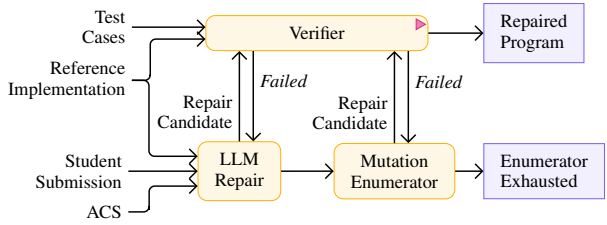

The repair module takes the \acf{ACS} produced by the Fault Localizer and tries to modify it to correct the original program.
It is possible that the \ac{ACS} produced by the fault localizer is empty. This happens when there are no incorrect rules (only missing rules) or when the fault localizer is unable to identify them.
If the \ac{ACS} is empty, \formhe attempts to extend the submitted program with new lines while maintaining the original ones.

\Cref{fig:repair} shows an overview of the Repair module. \formhe possesses two ways to repair programs: a fine-tuned Large Language Model and a program mutation enumerator.
First, \formhe attempts to repair the program using the fine-tuned LLM. This is done in a feedback loop where we attempt to refine the answers provided by the LLM until we find a correct repair. After a preset number of iterations of this loop, if no repair has been found, \formhe falls back to the mutation-based repair.
In this second phase, the Mutation Enumerator takes the original student submission and the \ac{ACS} identified by the fault localizer, and generates mutations of the identified lines. As in the LLM Repair, this starts a feedback loop where different repair candidates are enumerated until we either find one that is correct or the enumerator runs out of possible mutations.

The verification process is similar to assessing a student's submission for correctness. \change{We replace the \ac{ACS} in the student submission with the repair candidate and check if all answer sets of the resulting program are answer sets of the reference implementation for all test cases.} We also ensure that the program generates at least one answer set for each input, mirroring the behavior of the reference implementation.
When \formhe identifies a correction, it offers the student a hint in the form of a program with holes where changes were introduced to produce a repair. This hint is more precise than the \ac{ACS} produced by the fault localizer.

\subsection{LLM Program Repair}
Our LLM repair approach consists of using a fine-tuned LLM to obtain candidate repairs \change{of the \ac{ACS} found during fault localization}. The model was fine-tuned on synthetic data using an input prompt containing the name of the problem, the reference implementation, the student submission and the set of faulty lines (e.g., the ACS).
As introduced in the Fault Localization section, similarity-based methods can benefit from having a more closely related correct implementation instead of the reference one. 
Hence, in the LLM-based repair, we replace the reference implementation in the input prompt whenever a closer correct implementation is available.

If the repair candidate produced by the LLM is incorrect, \formhe enters a feedback loop where it runs fault localization on that candidate and then tries to further repair it with the LLM again. We do this for a preset number of times (3 by default), and if no correction is found, we fall to the mutation-based repair approach.

\subsection{Mutation Program Repair} \label{sec:enumeration}

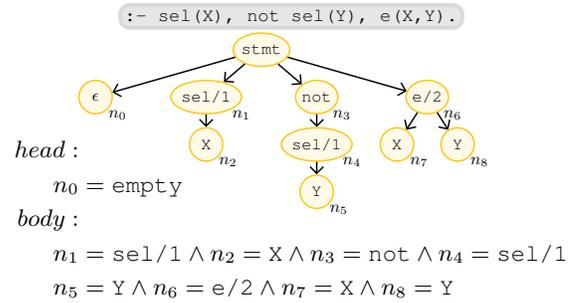
\begin{figure}[tb]
    \small
        \centering
        \aspinline{\scriptsize :- sel(X), not sel(Y), e(X,Y).}
        \scalebox{0.7}{\begin{tikzpicture}
    [thick,thing/.style={draw=c4,fill=c4!10,inner xsep=0.9em,inner ysep=0.6em},
        component/.style={draw=c2,fill=c2!10,inner xsep=0.9em,inner ysep=0.6em,rounded corners=0.5em},
        grow via three points={one child at (0,-0.9) and two children at (-1.05,-0.9) and (1.05,-0.9)},
        level/.style={arrows = {-Straight Barb[length=0.4em]},thick},
        tnode/.style={ellipse,inner xsep=0em,inner ysep=0em,font=\ttfamily,minimum height=2em,minimum width=2em},
        filled/.style={tnode,draw=c1,fill=c1!10},
        blank/.style={tnode,draw=black}]

    \node[filled] (root-1) {stmt}
    child {node[filled] (n0) {\(\epsilon\)}}
    child {node[filled] (n1) {sel/1}
            child {node[filled] (n2) {X}}
        }
    child {node[filled] (n3) {not}
            child {node[filled] (n4) {sel/1}
                    child {node[filled] (n5) {Y}}
                }
        }
    child {node[filled] (n6) {e/2}
            child {node[filled,xshift=1.5em] (n7) {X}}
            child {node[filled,xshift=-1.5em] (n8) {Y}}
        };

    \begin{scope}[nodes = {below right = -3pt and -3pt}]
        \foreach \x in {0,...,8}{%
                \node at (n\x.south east) {$n_\x$};
            }
    \end{scope}

\end{tikzpicture}}
        \vspace{-4.5em}
        \begin{align*}
            head:                                                                  \\
            n_0 & = \texttt{empty}                                                 \\
            body:                                                                  \\
            n_1 & = \texttt{sel/1} \land n_2 = \texttt{X} \land n_3 = \texttt{not} \land n_4 = \texttt{sel/1} \\
            n_5 & = \texttt{Y} \land n_6 = \texttt{e/2} \land n_7 = \texttt{X} \land n_8 = \texttt{Y}
        \end{align*}
    \caption{An \ac{ACS} with a single statement, its \acs*{AST} representation, and the \acs*{SMT} encoding for that tree.}
    \label{fig:mutation-ex}
\end{figure}

The second repair technique finds repair candidates by enumerating mutations of the \change{\ac{ACS}}. The first step in this process is to convert the \ac{ACS} into \formhe's \acl{DSL}\change{, which contains the most common \ac{ASP} operators,} and create an \ac{AST}. (More details on \formhe's \acs{DSL} can be found in the appendix.) Then, the \ac{AST} is encoded into a \ac{SMT} formula~\cite{DBLP:journals/pvldb/MartinsCCFD19,DBLP:journals/pvldb/OrvalhoTVMM20,DBLP:conf/kbse/RamosPLMM20,DBLP:conf/icse/NiRYLMMG21,DBLP:conf/fase/BrancasTVMM24,ferreira2024reverse}. \Cref{fig:mutation-ex} shows an example of an \ac{ACS} and respective \ac{AST} and \ac{SMT} encoding. Relaxing each of the equalities in this formula allows us to use an \ac{SMT} solver to enumerate program mutations. For each node \(n_i\) we introduce a relaxation variable \(r_i\). For example, relaxing \(n_2 = \texttt{X}\) would become \((n_2 = \texttt{X} \lor r_2)\) and allow us to enumerate \aspinline{:- sel(\textcolor{c4}{Y}), not sel(Y), e(X,Y).}.

\label{sec:pruning}
To reduce the number of enumerated programs that are either semantically incorrect ASP programs or that are mathematically equivalent, we enforce several restrictions that prune enumerated programs. These include symmetry breaking for commutative operations, ignoring neutral and absorbing elements for addition and multiplication, and several other optimizations related to ASP semantics, such as pruning invalid constructions of rule heads.

\section{Evaluation} \label{sec:results}

To properly evaluate \formhe's capabilities, we collected programs from students taking an university class on Automated Reasoning. Over two years, we collected 115 programs submitted by students on 5 different assignments. In these assignments, students encoded classical graph problems such as graph coloring and vertex cover, as well as set problems such as pairwise disjoint sets and set cover.
Of those 115 instances, we received 63 correct submissions and 52 semantically incorrect submissions.

We also created synthetic benchmarks in order to better evaluate our tool. These benchmarks were created by randomly introducing 1 to 8 mutations in the correct submissions, including some of the most common bug types, such as using a wrong predicate name or forgetting a constraint. We created 95k such instances for training the different machine learning models and 500 instances for evaluation.

This section answers the following research questions:
\begin{enumerate}
  \item[\textbf{Q1.}] How effective are the fault localization approaches?
  \item[\textbf{Q2.}] Can we improve the fault localization through a combination of approaches?
  \item[\textbf{Q3.}] How effective is the program repair?
\end{enumerate}

We evaluated our tool using an Intel Xeon Silver 4210R and imposed a limit of 10 minutes (wall clock time) and 60GB of RAM per instance. Limits were strictly imposed using Runsolver~\cite{runsolver}.
\formhe is implemented in Python and uses the Clingo ASP grounder and solver \cite{DBLP:journals/tplp/GebserKKS19} version 5.6.2. For the enumeration of program mutations, \formhe uses a modified version of the Trinity framework \cite{DBLP:journals/pvldb/MartinsCCFD19}.
Finetuning and evaluation of the different \acp{LLM} was performed using 5x Nvidia RTX A4000. \formhe's source code, data and logs are available as supplemental material.

\begin{table*}[tb]
\centering
\small
\scalebox{0.90}{
\begin{tabular}{l|cr@{\hspace{3\tabcolsep}}rc|r@{\hspace{0.3\tabcolsep}}rr@{\hspace{0.3\tabcolsep}}rr@{\hspace{0.3\tabcolsep}}rr@{\hspace{0.3\tabcolsep}}rr@{\hspace{0.3\tabcolsep}}r}
  \toprule
 & \multicolumn{4}{c|}{Exact + Superset + Some} & \multicolumn{2}{c}{Exact Faults} & \multicolumn{2}{c}{Superset Faults} & \multicolumn{2}{c}{Some Faults} & \multicolumn{2}{c}{Fault Not} & \multicolumn{2}{c}{Wrong} \\ 
 & \multicolumn{4}{c|}{Faults Identified} & \multicolumn{2}{c}{Identified} & \multicolumn{2}{c}{Identified} & \multicolumn{2}{c}{Identified} & \multicolumn{2}{c}{Identified} & \multicolumn{2}{c}{Identification} \\ 
 & & \scriptsize \color{gray} Real & \scriptsize \color{gray} Synth. & & \scriptsize \color{gray} Real & \scriptsize \color{gray} Synth. & \scriptsize \color{gray} Real & \scriptsize \color{gray} Synth. & \scriptsize \color{gray} Real & \scriptsize \color{gray} Synth. & \scriptsize \color{gray} Real & \scriptsize \color{gray} Synth. & \scriptsize \color{gray} Real & \scriptsize \color{gray} Synth.  \\ 
  \midrule
MSICS & & 63.5\% & 46.0\% & & 40.4\% & 24.2\% & 0.0\% & 0.4\% & 23.1\% & 21.4\% & 23.1\% & 42.6\% & 13.5\% & 11.4\% \\ 
   \midrule
Line Matching & & 69.2\% & 48.2\% & & 34.6\% & 20.2\% & 11.5\% & 3.8\% & 23.1\% & 24.2\% & 25.0\% & 40.4\% & 5.8\% & 11.4\% \\ 
   \midrule
Gemma 2B & & 90.4\% & \textbf{100.0\%} & & 59.6\% & 99.6\% & 17.3\% & 0.0\% & 13.5\% & 0.4\% & 5.8\% & 0.0\% & 3.8\% & 0.0\% \\ 
  CodeGemma 2B & & 84.6\% & \textbf{100.0\%} & & 50.0\% & 99.2\% & 19.2\% & 0.0\% & 15.4\% & 0.8\% & 9.6\% & 0.0\% & 5.8\% & 0.0\% \\ 
  StarCoder2 3B & & 65.4\% & 99.2\% & & 26.9\% & 96.6\% & 7.7\% & 0.2\% & 30.8\% & 2.4\% & 26.9\% & 0.6\% & 7.7\% & 0.2\% \\ 
    Phi 3 mini & & 84.6\% & \textbf{100.0\%} & & 51.9\% & 100.0\% & 13.5\% & 0.0\% & 19.2\% & 0.0\% & 9.6\% & 0.0\% & 5.8\% & 0.0\% \\ 
   \midrule
\formhe w/o LLM & & 82.7\% & 69.8\% & & 44.2\% & 32.2\% & 13.5\% & 8.8\% & 25.0\% & 28.8\% & 11.5\% & 19.8\% & 5.8\% & 10.4\% \\ 
  \hspace{-1.32em}$\rightarrow$ \textbf{\formhe} (with Gemma 2B) & & \textbf{94.2\%} & 97.8\% & & 61.5\% & 77.6\% & 23.1\% & 17.8\% & 9.6\% & 2.4\% & 1.9\% & 0.0\% & 3.8\% & 2.2\% \\ 
\formhe w/o Impl. Choosing & & 92.4\% & 98.0\% & & 63.5\% & 71.4\% & 21.2\% & 25.8\% & 7.7\% & 0.8\% & 1.9\% & 0.0\% & 5.8\% & 2.0\% \\ 
   \midrule
\change{DWASP\textsuperscript{\textdagger}} & & 69.3\% & 31.0\% & & 46.2\% & 19.6\% & 1.9\% & 0.6\% & 21.2\% & 10.8\% & 30.8\% & 62.8\% & 0.0\% & 6.2\% \\ \bottomrule
  \end{tabular}
  }
  \caption{Results for different fault localization methods for real and synthetic instances. Label meanings: ``Exact Faults Identified'' --- the method identified all the faulty lines and no others; ``Superset Faults Identified'' --- the method identified all the faulty lines, but also some others; ``Some Faults Identified'' --- the method identified some of the faults, but not all; ``Faults Not Identified'' --- the program had faulty lines but the method was unable to identify them; ``Wrong Identification'' --- the method only identified lines as faulty that were actually correct. \change{The first column shows the sum of the Exact, Superset and Some Faults Identified columns and represents the percentage of instances where at least one of the faults present in the program was found.}} 
\label{tab:fault-localizer}
\end{table*}

\subsection*{Q1: How effective are the different fault localization approaches?}

This section explores the different fault localization approaches in \formhe. Note that while the different fault localization approaches are intended to be used together, they can also be used in isolation.
\Cref{tab:fault-localizer} shows the percentage of instances where each approach correctly identified the faults for real and synthetic instances.

\paragraph{\acs*{MSICS} Method}
The \ac{MSICS} fault localization method has the largest number of Exact Faults Identified out of the two \change{traditional} methods.
However, due to the non-monotonicity of \ac{ASP}, this method produces some unexpected results, with a large number of non-identified faults and wrong identifications. For instance, when the body of a rule has a bug and is never satisfied, that rule does not contribute to the logical behavior of the program and will thus not be identified by this method. Even so, this method correctly identifies at least one fault in a large number of instances (64\%) and provides a good baseline.

\paragraph{Line Matching}
The line matching algorithm exploits similarities that may exist between the student submission and the reference implementation. While this approach may be less helpful for complex programs where there are many ways to solve the problem, for simple instances such as the ones used in introductory \ac{ASP} classes, it performs well, since most solutions will follow similar patterns.
The large number of wrong identifications for synthetic instances is due to these instances using a normalized representation which can sometimes cause false positives with the reference implementation (for example, \aspinline{0 \{\dots\} 1} is syntactically different but equivalent to \aspinline{\{\dots\} 1}).

\paragraph{Large Language Models}
We use open-access \ac{LLM}s with a small number of parameters for three reasons: (1) closed-access models are prohibitive due to cost and students' data privacy concerns, (2) models with a large number of parameters require large amounts of compute power and take longer to produce answers, which is undesirable in a classroom, and (3) since we transform the models into classifiers with few outputs, too many parameters can be detrimental and lead to overfitting. Furthermore, we use fine-tuned models due to two major issues: (1) many models have trouble reasoning about \ac{ASP}, likely due to a small proportion of this language in their training sets, and (2) some models have trouble respecting specific output formats which makes it hard to automatically extract the relevant parts of the answer for verification and/or use in other \formhe modules. These issues can be partially alleviated by using large model sizes, but, as explained, that is undesirable for \formhe's intended use.

We fine-tuned four state-of-the-art models to evaluate our \ac{LLM} fault localization approach.
We chose two models with 2B parameters: Gemma~\cite{DBLP:journals/corr/abs-2403-08295} and CodeGemma~\cite{codegemmateam2024codegemmaopencodemodels}, a model with 3B: StarCoder 2~\cite{DBLP:journals/corr/abs-2402-19173}, and a model with 4B: Phi 3~\cite{DBLP:journals/corr/abs-2404-14219}.
These models were fine-tuned using 95k synthetically generated incorrect programs. We used \acl{PEFT}~\cite{peft} (in particular \ac{LoRa}~\cite{DBLP:conf/iclr/HuSWALWWC22}) to decrease the memory requirements during training.

\Cref{tab:fault-localizer} shows the results for different models for real and synthetic instances. Note that the synthetic results are for the 500 evaluation instances and not for the 95k used for training. 
Of the four models, Gemma performed the best overall. While Phi 3 performs better than Gemma for synthetic instances, the same is not true for real instances. This indicates that Phi 3 has likely overfitted the training data.
Furthermore, the overall disparity between the results for synthetic and real instances is expected since the models were also fine-tuned using synthetic instances.
Even with these caveats, the results for real instances are generally better than the other fault localization approaches.

\paragraph{Related Methods}
To compare \formhe with previous approaches, we also implemented the core DWASP~\cite{DBLP:journals/tplp/DodaroGRRS19} fault localization method \change{(excluding the interactive portion)} in our tool. The results for this method can be found as DWASP\textsuperscript{\textdagger} in \Cref{tab:fault-localizer}. The DWASP\textsuperscript{\textdagger} approach is very similar to the \ac{MSICS} method, and this shows in the results, with comparable performance for these approaches. Even though the DWASP\textsuperscript{\textdagger} slightly outperforms the \ac{MSICS} method in real instances, replacing it in \formhe's default configuration results in worse performance, suggesting that DWASP\textsuperscript{\textdagger} has a greater overlap with the other methods.

\subsection*{Q2: Can we improve the fault localization through a combination of approaches?} \label{sec:results-fl-combined}

\Cref{tab:fault-localizer} shows the results for a configuration of \formhe without using \acp{LLM} (``\formhe w/o LLM'') and for the recommended configuration using the Gemma 2B model (``\formhe (with Gemma 2B)''). Combining the \ac{MSICS} and Line Matching methods provides a big improvement in performance compared with using the isolated approaches. Furthermore, in deployment scenarios capable of using deep learning, adding the \ac{LLM} fault localization method provides a further boost, with the recommended configuration finding all faults in \(85\%\) of real submissions and at least one of the faulty lines in \(94\%\). Even though \formhe has a much larger number of ``Superset Faults Identified'' cases than other approaches, for 78\% of those cases only 1 extra line was identified. This means that \formhe is still helping the student focus on the problematic section of the program.

We also explored the impact of using the closest correct implementation for the Line Matching and LLM methods instead of the reference implementation. For a given instance, we only consider previously submitted correct solutions, simulating real-time usage of \formhe. The line labeled ``\formhe w/o Impl. Choosing'' shows the effects of disabling this feature compared with the default configuration of \formhe. Disabling the feature has a small impact, decreasing the number of instances in which we find at least one fault and raising the number of Wrong Identifications.

Our combined approach can successfully incorporate the different fault localization methods, obtaining the best overall results of all techniques for real instances. \formhe can provide students actionable feedback for the majority of faulty submissions. Furthermore, even for submissions where we cannot identify the fault, we can still inform the student that the program is incorrect and provide failing test cases with missing and extra answer sets.

\subsection*{Q3: How effective is the program repair?}
\label{sec:asp-repair}

\begin{figure}[tb]
  \centering
  \scalebox{1}{\includegraphics{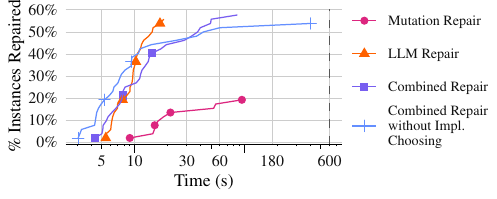}}
  \caption{Percentage of submissions (i.e., real instances) repaired at each point in time.}
  \label{fig:repair-cactus}
\end{figure}

\change{This section explores the performance of the repair module using the best-performing fault localization method (\formhe with Gemma 2B).}
\Cref{fig:repair-cactus} shows how many real instances can be repaired under \(x\) seconds for different configurations of \formhe. Times shown in this plot include the time for the fault localizer (5 seconds on average). The model used for the \ac{LLM} and Combined configurations was CodeGemma 7B finetuned using \ac{LoRa} and 4 bit quantization. \change{We used a larger number of parameters for repair than for fault localization since the repair task is more complex.}

The ``Mutation Repair'' and ``LLM Repair lines'' refer to using each of the two repair techniques in isolation. The LLM approach has a much greater repair rate at \(56\%\) compared to \(19\%\) for the mutation-based. This is partially due to the mutation-based method having been tuned for harder instances to try to repair instances where the LLM fails.

The combined approach slightly improves over using just the \ac{LLM} repair, with \(58\%\) repaired instances. Even though mutation-based repair makes a small contribution to the combined approach, it is not dependent on having synthetic instances and performing computationally intensive fine-tuning of models. It can thus be used in low-resource situations and provides a safe fallback when the \ac{LLM} fails.
Observe that \formhe repairs almost all instances in 1 minute or less. This is ideal for classroom situations since it is not desirable for students to wait a long time for feedback.

The Combined Repair configuration uses the closest correct implementation available in the \ac{LLM} input prompt. The impact of disabling this feature is shown in \cref{fig:repair-cactus} as ``Combined Repair without Impl. Choosing''. This feature has a larger impact on repair performance than on fault localization with a 4pp. drop in repair rate when it is disabled.
There is also a large impact on the repair time. Using the closest solution takes at most about 2 minutes versus 8 minutes when using the original reference solution.

Although not shown in the figure, synthetic instances show the same behavior as in fault localization: the \ac{LLM} approach performs much better at synthetic instances than real instances (\(97\%\) vs \(56\%\)), while the traditional method (mutation-based) performs similarly on the two types of instances (\(20\%\) vs \(19\%\)). This can be explained by the LLM having been trained on synthetic instances (although not the same ones as used for evaluation).
We also finetuned other models for the repair task besides CodeGemma 7B. These are: 8 bit quantized CodeQwen 1.5 7B~\cite{codeqwen} with \(48\%\) combined repair rate for real instances, Phi 3 mini with \(40\%\) and CodeGemma 2B with \(37\%\).

\formhe's repair rate depends greatly on the fault localization outcome. Looking at all 552 instances (real + synthetic), \formhe can repair 96\% of the ones with Exact Faults Identified, 92\% of Superset Faults Identified and 71\% of Some Faults Identified. This shows that better fault localization leads to better repair rates and also that it is preferable to identify a superset of the faulty lines than to only identify some of the faults. Furthermore, \formhe can also repair  50\% of the instances where the faults were Not Identified or Wrongly Identified. 
This occurs in instances where the faults are missing ASP rules. Although the fault localization module does not always detect this, the repair module can sometimes solve the problem by synthesizing new rules.

\section{Related Work} \label{sec:related-work}

\subsubsection{Fault Localization}

Work on fault localization for ASP has been ongoing.
SeaLion \cite{DBLP:journals/tplp/OetschPT18} is an \ac{ASP} debugger which allows users to compute answer sets step-by-step. Initially, no rules are active and the answer set is empty. Then, at each step, the user can select a new rule to be active, producing changes in the current answer set. This allows a user to find out at which point the answer set deviates from what is expected. This tool requires some intuition to be used effectively, since users need to choose the right rules at each step and know when the computation has ``gone wrong''.
DWASP \cite{DBLP:journals/tplp/DodaroGRRS19} is a debugger based on the WASP solver \cite{DBLP:conf/lpnmr/AlvianoDLR15} that requires an input (e.g., an example graph for the vertex cover problem), and a test case (a set of atoms that should be part of some answer set but are not). DWASP then computes a minimal set of lines that make the test case incoherent with the input. Although the \emph{reason of incoherence} is minimal from a logic standpoint, it usually includes lines that are not faulty. As such, DWASP implements an interactive debugger that allows users to refine the reason for incoherence.
Although this approach is more automated than previous work, it still requires users to (1) select a good input and test case that expose the fault and (2) refine the set of possible lines by answering questions that require complex reasoning.

Recently, many approaches to fault localization using deep learning have been proposed.
DeepFL~\cite{DBLP:conf/issta/LiLZZ19} uses a deep learning model to combine the results of many fault localization techniques (such as mutation-based, textual-similarity, among others) into a single suspiciousness score for each line.
TRANSFER~\cite{DBLP:conf/icse/MengW00022} uses bidirectional LSTM-based classifiers to compute deep semantic features of the buggy programs. Then it uses a different model to combine these features with spectrum and mutation-based metrics and produce suspiciousness scores for each line of the program.
LLMAO~\cite{DBLP:conf/icse/YangGMH24} uses a two-step approach where a pretrained \ac{LLM} is used to obtain a representation for each line in the program (the model's hidden state), and then a bidirectional transformer model is used to transform the sequence of representations into a suspiciousness score for each line. This two-step approach is used to avoid the \ac{LLM} having to ``remember'' the full program in the hidden state of the last token, which could make it difficult to use with very large programs.

\subsubsection{Program Repair}

\ac{APR} encompasses techniques aimed at automatically fixing bugs in programs, spanning from imperative~\cite{DBLP:journals/tosem/RamosLMMG24,DBLP:journals/tse/GouesNFW12} to declarative paradigms~\cite{DBLP:conf/kbse/WangSK18,DBLP:conf/kbse/BridaRZBNAF22,DBLP:conf/issta/ZhengNBRAFB22}. To the best of our knowledge, \formhe is the first automatic program repair tool for \ac{ASP}. However, work has been ongoing in automatically repairing other declarative languages such as Alloy.
ARepair \cite{DBLP:conf/kbse/WangSK18} and ICEBAR \cite{DBLP:conf/kbse/BridaRZBNAF22} are two repair tools for Alloy that use AlloyFL as a fault localizer. ARepair uses a sketch-based approach: based on the suspiciousness scores computed by AlloyFL, ARepair selects suspicious nodes and replaces them with holes. Then a synthesizer tries to fill those holes to produce a correct program.
ICEBAR builds upon ARepair by introducing an extra form of specification: a property-based oracle that validates if the model respects some property. Using this oracle, ICEBAR iteratively increases the set of test cases used by ARepair, improving the quality of the repairs and reducing overfitting.
ATR \cite{DBLP:conf/issta/ZhengNBRAFB22} is a repair tool for Alloy that uses FLACK as a fault localizer. ATR uses pairs of counterexamples and closely related satisfying instances. Based on these pairs, ATR constructs
candidate repairs in a bottom-up manner. Finally, ATR tries to replace the suspicious statements produced by FLACK with these candidates.

\section{Conclusion} \label{sec:conclusions}

This paper proposes \formhe, a fault localization and program repair tool for ASP that combines logic techniques with machine learning. \formhe helps students who are using declarative languages for the first time and do not have the intuition necessary to use other forms of debugging. \formhe assists them in finding and correcting faults in their programs.
\change{\formhe can offer valuable insights to students by correctly identifying faults in 94\% of incorrect submissions and providing repair hints in 56\% of cases. Furthermore, even when the fault localization and repair are unsuccessful, \formhe can still provide students a failing example that points them in the right direction.}
\formhe with only logic-based techniques has been used successfully for two years in an Automated Reasoning class where we collected the student submissions. 
In next year's classes, we will include all techniques proposed in the paper.

\bibliography{aaai25}

\clearpage

\appendix

\section{Technical Appendix}

\subsection{\acs*{MSICS} Program Relaxation}

\begin{figure}[h]
    \centering
    \begin{subfigure}[t]{\linewidth}
        \small
        \begin{asplisting}[minipage]
v(X) :- e(X,_).
v(X) :- e(_,X).
k { sel(X) : v(X) } k.
:- not sel(X), not sel(Y), e(X,Y).
        \end{asplisting}
        \caption{Student submission.} \label{fig:mcs-code-student}
    \end{subfigure}
    \begin{subfigure}[t]{\linewidth}
        \small
        \begin{asplisting}[minipage,colback=c2!10!white,colframe=c2!30!white,enhanced,remember as=soft]
v(X) :- e(X,_).
v(X) :- e(_,X). %*\tikz[remember picture] \node [] (relax-line3){};*)
k { sel(X) : v(X) } k. %*\tikz[remember picture] \node [] (relax-line4){};*)
:- not sel(X), not sel(Y), e(X,Y).
        \end{asplisting}
        \vspace{-1.2em}%
        \begin{asplisting}[minipage,colback=c34!10!white,colframe=c34!30!white,listing options={numbers=left,
      basicstyle=\small\ttfamily,
      numberstyle=\color{black}\scriptsize,
      showstringspaces=false,
      breaklines=true,
      escapeinside={\%*}{*)},
      language=prolog,
      aboveskip=\smallskipamount,firstnumber=5,
      backgroundcolor={}},enhanced,remember as=hard]
%*\textcolor{c3}{\#const k = 3.}\tikz[remember picture] \node [] (relax-line5){};*)
%*\textcolor{c3}{e(1,2).}\tikz[remember picture] \node [] (relax-line6){};*)
%*\textcolor{c4}{sel(2).}\tikz[remember picture] \node [] (relax-line7){};*)
        \end{asplisting}
        \nointerlineskip
        \begin{tikzpicture}[remember picture, overlay]
            \node[] [below left=0.5em of soft.north east] {\normalsize\textbf{\textcolor{c2}{Soft}}};
            \node[] [below left=0.5em of hard.north east] {\normalsize\textbf{\textcolor{c34}{Hard}}};
            \node[circle,fill=c2,inner sep=1pt,text=white] [right=2em of $(relax-line3)!0.5!(relax-line4)$] {\normalsize1};
            \node[circle,fill=c3,inner sep=1pt,text=white] [right=2em of $(relax-line5)!0.5!(relax-line6)$] {\normalsize2};
            \node[circle,fill=c4,inner sep=1pt,text=white] [right=0.4em of relax-line7] {\normalsize3};
        \end{tikzpicture}
        \caption{Relaxed program.} \label{fig:mcs-code-relaxed}
    \end{subfigure}
    \caption{Student submission for the vertex cover problem before and after relaxation.}
    \label{fig:mcs-code}
\end{figure}

\Cref{fig:mcs-code} shows how to relax the student submission for the example in \cref{sec:fault-localization}. First, we turn each of the lines in the student program into soft constraints \tikzcircleone. Next, we specify the test case \tikzcircletwo, and the missing answer set \tikzcirclethree as hard constraints. From these soft and hard constraints, we can use an \ac{MSICS} algorithm \cite{DBLP:conf/sat/Mencia020} to compute a minimal set of lines of the program \tikzcircleone that need to be removed or modified such that the missing answer set \tikzcirclethree can become a solution for the test case \tikzcircletwo.

\subsection{\formhe's \acs*{DSL}}

Like most enumeration-based program synthesis tools, \formhe's synthesizer uses a \ac{DSL} to define the repair search space. \formhe's DSL includes the most common ASP operators, such as Boolean operators and aggregate rules. Furthermore, it also takes into account information from the student submission, such as predicate and variable names.

\begin{figure}[tb]
    \centering
    \begin{rcfg}
      & stmt \produces \prodtwo{stmt}{head ?}{atom \ast} \\
      & head \produces \prodfour{aggregate}{term ?}{atom}{atom}{term ?} \\
      \alignprod \ror atom \\
      & atom \produces predicate\texttt{(}pred\_term \ast\texttt{)} \ror \prodone{not}{atom}\\
      \alignprod \ror \prodone{classical\_not}{atom}\\
      \alignprod \ror \prodtwo{eq}{term}{term} \ror \prodtwo{neq}{term}{term}\\
      \alignprod \ror \prodtwo{lt}{term}{term} \ror \prodtwo{le}{term}{term}\\
      \alignprod \ror \prodtwo{gt}{term}{term} \ror \prodtwo{ge}{term}{term}\\
      \alignprod \ror \prodtwo{pool}{atom}{atom}\\
      & pred\_term \produces term \ror interval \ror \texttt{\_}\\
      & term \produces 0 \ror 1 \ror constant \ror variable \\
      \alignprod \ror \prodtwo{add}{term}{term} \ror \prodtwo{sub}{term}{term} \\
      \alignprod  \ror \prodtwo{mul}{term}{term} \ror \prodtwo{div}{term}{term}\\
      \alignprod  \ror \prodone{abs}{term}\\
      & interval \produces \prodtwo{interval}{term}{term}
    \end{rcfg}
    \caption{\formhe's base \acl*{DSL}.}
    \label{fig:dsl}
\end{figure}

\Cref{fig:dsl} shows the base \ac{DSL} supported by FormHe. The three productions not defined in this figure (\(predicate\), \(constant\), and \(variable\)) are dependent on the user submission and on the \ac{ACS}.
The \(predicate\) production can be any predicate used in the student submission, as well as any of the solution predicates for that instance. \formhe ensures that the arity of each predicate is respected. Likewise, \(constant\) can be any constant used in the user submission. Finally, \(variable\) can be any of the variables used in the \ac{ACS} as well as completely new variables.

\subsection{Prompt Templates \& Training Methods}

\begin{figure}[tb]
    \centering
    \begin{subfigure}[t]{\linewidth}
    \begin{tcblisting}{listing only,enhanced,listing options={%
        basicstyle=\normalfont\ttfamily\footnotesize,
        showstringspaces=false,
        breaklines=false,
        escapeinside={\%*}{*)},
        breakautoindent=false,
        breakindent=0ex, postbreak=\usebox\mypostbreak
    },halign=justify,top=-0.4em,right=0.1em,left=0.1em,bottom=-0.4em}
<|problem|>%*\textbf{\{Problem Title\}}*)
<|reference_program|>%*\textbf{\{Reference Program\}}*)
<|incorrect_program|>%*\textbf{\{Incorrect Program\}}*)
\end{tcblisting}
    \caption{Fault localization prompt.}
    \label{fig:prompt-fl}
    \end{subfigure}
    \begin{subfigure}[t]{\linewidth}
    \begin{tcblisting}{listing only,enhanced,listing options={%
        basicstyle=\normalfont\ttfamily\footnotesize,
        showstringspaces=false,
        breaklines=false,
        escapeinside={\%*}{*)},
        breakautoindent=false,
        breakindent=0ex, postbreak=\usebox\mypostbreak
    },halign=justify,top=-0.4em,right=0.1em,left=0.1em,bottom=-0.4em}
<|problem|>%*\textbf{\{Problem Title\}}*)
<|reference_program|>%*\textbf{\{Reference Program\}}*)
<|incorrect_program|>%*\textbf{\{Incorrect Program\}}*)
<|fl|>%*\textbf{\{ACS Lines\}}*)
<|missing_lines|>%*\textbf{\{Yes} \textit{or} \textbf{No\}}*)
<|correction|>
\end{tcblisting}
    \caption{Repair prompt.}
    \label{fig:promp-repair}
    \end{subfigure}
    \caption{\ac{LLM} input prompts for the fault localization and repair modules.}
    \label{fig:prompts}
\end{figure}

Figures~\ref{fig:prompt-fl} and~\ref{fig:promp-repair} show the prompts used for finetuning the fault localization and repair models, respectively. In these prompts, the strings \texttt{<|...|>} each refer to a custom token added to the models' embeddings. This improves ease of training and decreases the LLM's response time (by decreasing the total number of tokens). \change{For the finetuning of our models, we used 4 epochs, a learning rate of \(10^{-4}\), batch size of 1 to 8 (depending on VRAM requirements),  gradient accumulation, a LoRA \(r\) value of 8, LoRA alpha of 8 and LoRA dropout of 0.05.}

The fault localization model was trained using Multi Label Sequence Classification, where each label X from 1 to \texttt{MAX\_LINES} represents if line number X is faulty or not. Although not mentioned in the main paper, we also included an extra label representing if the program has missing lines or not.

The repair model was trained using Supervised Finetuning. Besides the inputs mentioned in the main paper, the repair prompt also receives information about if the program has missing lines or not. This information comes directly from the extra label in the fault localization model.

\end{document}